\documentclass{kluwer}    
\usepackage{graphicx}
\newdisplay{guess}{Conjecture}

\begin{document}                                                                                   
\begin{article}
\begin{opening}         
\title{AGNs and microquasars as high energy $\gamma$-ray sources} 
\author{Josep M. \surname{Paredes}}  
\runningauthor{Josep M. Paredes}
\runningtitle{AGNs and microquasars}
\institute{Departament d'Astronomia i Meteorologia, Universitat de
Barcelona, Av. Diagonal 647, 08028 Barcelona, Spain}
\date{September 10, 2004}

\begin{abstract}
The extragalactic analogs of the microquasars, the quasars, are strong
$\gamma$-ray emitters at GeV energies. It is expected that
microquasars are also $\gamma$-ray sources, because of the analogy
with quasars and because theoretical models predict the high-energy
emission. There are two microquasars that appear as the possible
counterparts for two unidentified high-energy $\gamma$-ray sources.
\end{abstract}
\keywords{X-rays: binaries - stars: individual: LS~5039, LS~I~+61~303 -
gamma-rays: observations}

\end{opening}           

\section{Introduction}  
The microquasar phenomenon has grown in prominence recently, as it has
been found that most X-ray binaries show radio emission, associated
with jets - collimated beams of relativistic plasma.  The ejection takes
place in a bipolar way perpendicular to the accretion disk associated
with the compact star, a black hole or a neutron star. The word
microquasar itself was chosen by the analogy of these astronomical
objects with quasars and other active galactic nuclei (AGNs) at
cosmological distances (Mirabel \& Rodr\'{\i}guez \citeauthor{mir99}).
The analogy quasar-microquasar goes beyond a simple morphological
resemblance. Today, there is growing evidence to think that the
physics involved in both types of objects is the same, or at least
very similar. The key difference would be the distinct order of
magnitude of the most significant parameters, especially the mass of
the compact object.
 
\section{AGNs as $\gamma$-ray sources}
AGNs are extragalactic sources whose spectra extend from radio waves
to $\gamma$-rays. Thanks to the Compton Gamma Ray Observatory (CGRO)
it is now well established that AGNs are strong $\gamma$-ray
emitters. The Energetic Gamma Ray Experiment (EGRET) on board the CGRO
produced the 3rd EGRET catalogue (Hartman et al.~\citeauthor{har99})
that contains 271 sources detected at energies $>$ 100~MeV. Most of
them (about 168 sources) remain unidentified and 72 of these
unidentified sources are at absolute Galactic latitudes lower than
10$^{\circ}$.  The identified sources are mainly AGNs (Thomson et
al. 1995), and more AGNs are expected to be found among the still
unidentified sources with high Galactic latitude. All of these
detected AGNs are blazars; no radio-quiet AGNs has been
identified so far in the EGRET data. In fact this is not surprising
because blazars are able to generate high-energy particles that can
produce $\gamma$-rays via inverse Compton (IC) scattering, as well as
present relativistic beaming, which is important to avoid
photon-photon collision and amplify the flux.

More than two dozen jets have been detected at X-rays in AGNs, with
most of them being radio galaxies (both FRI and FRII). Although
synchrotron self Compton (SSC) models cannot generally explain the
level of X-ray emission, those models based on the IC scattering of
seed photons of the nucleus and the CMB radiation by the relativistic
electrons in the jet not only can better explain such X-ray levels but
also can produce efficiently high-energy $\gamma$-ray emission
(Tavecchio et al.~\citeauthor{tave00}, Celotti et
al.~\citeauthor{cel01}).

\section{X-ray binaries and microquasars}

The most recent catalogue of High Mass X-ray Binaries (HMXBs) contains
131 sources (Liu et~al.~\citeauthor{liu00}), while the catalogue of
Low Mass X-ray Binaries (LMXBs) amounts to 149 objects (Liu
et~al.~\citeauthor{liu01}). Considering both catalogues together,
there are a total of 43 radio emitting sources. Some of these sources,
those which we define as microquasars, show direct evidence for a
relativistic radio jet (Rib\'o \citeauthor{rib02}, Rib\'o
\citeauthor{rib04}), while many others show radio emission which is
also almost certainly associated with a jet (Fender 2004).


At the time of writing, a total of 15 microquasar systems have been
identified. The observational data of these
microquasars at energies from soft to very
high-energy $\gamma$-rays, are quoted  in Table~\ref{detections}.

\begin{table}[t!]
{\small
\caption[]{High energy emission from microquasars}
\label{detections} 
\begin{tabular}{@{}l@{\hspace{0.07cm}}c@{\hspace{0.05cm}}cc@{\hspace{0.05cm}}c@{\hspace{0.05cm}}c@{\hspace{0.05cm}}c@{\hspace{0.05cm}}c@{\hspace{0.05cm}}c@{}}
             &                 &           &      &              &              &                           \\ 
\hline \noalign{\smallskip}
Name  & {INTEGRAL$^{\rm (a)}$}   & {BATSE$^{\rm (b)}$} 
&     COMPTEL$^{\rm (c)}$ & EGRET$^{\rm (d)}$ & Others$^{\rm (e)}$  \\ & 40$-$100 keV    &   160$-$430 keV & 1$-$30 MeV &  $>$ 100 MeV &  & \\ 
 & (count/s)  & (mCrab) & (GRO) & (3EG)\\
\noalign{\smallskip} \hline \noalign{\smallskip}
\multicolumn{6}{c}{\bf High Mass X-ray Binaries (HMXB)}\\
\noalign{\smallskip} \hline \noalign{\smallskip}

{\bf LS~I~+61~303} &  $-$  & 5.1$\pm$2.1  & J0241+6119? & J0241+6103? &  \\

{\bf V4641~Sgr}  & $-$ & $-$   & $-$ & $-$ &  \\

{\bf LS~5039} &  $-$ &  3.7$\pm$1.8 & J1823$-$12?  & J1824$-$1514? & \\
  
{\bf SS~433} & $<$1.02  &  0.0$\pm$2.8 & $-$ & $-$ &   \\
  
{\bf Cygnus~X-1}  & 66.4$\pm$0.1  & 924.5$\pm$2.5 &  yes &$-$& S\\
  
{\bf Cygnus~X-3}    &  5.7$\pm$0.1     & 15.5$\pm$2.1 & $-$   & $-$ & O, T?  \\

\noalign{\smallskip} \hline \noalign{\smallskip}
\multicolumn{6}{c}{\bf Low Mass X-ray Binaries (LMXB)}\\
\noalign{\smallskip} \hline \noalign{\smallskip}

      
{\bf Circinus~X-1}    &  $-$     &   0.3$\pm$2.6&  $-$   & $-$& \\
 
{\bf XTE~J1550$-$564} &  0.6$\pm$0.07    & $-$2.3$\pm$2.5 & $-$   & $-$&  \\
 
{\bf Scorpius~X-1}     &  2.3$\pm$0.1      &  9.9$\pm$2.2 &  $-$ & $-$ &   \\
  
{\bf GRO~J1655$-$40} & $-$    &   23.4$\pm$3.9 & $-$ & $-$& O    \\
   
{\bf GX~339$-$4}   &   0.55$\pm$0.03    &   580$\pm$3.5 & $-$&$-$& S    \\ 
  
{\bf 1E~1740.7$-$2942}&  4.32$\pm$0.03 &  61.2$\pm$3.7 & $-$& $-$& S  \\
 
{\bf XTE~J1748$-$288} &   $-$  &    $-$ & $-$ & $-$  &S        \\

{\bf GRS~1758$-$258}  &  3.92$\pm$0.03  &  38.0$\pm$3.0  & $-$ &$-$& S\\
   
{\bf GRS~1915+105}    &   8.63$\pm$0.13 &  33.5$\pm$2.7 & $-$  & $-$ & S, T?\\

\hline
\end{tabular}
{\small
Notes: $^{\rm (a)}$ The first IBIS/ISGRI soft gamma-ray galactic plane survey catalog 
(Bird et al.~\citeauthor{bir04}). $^{\rm (b)}$~BATSE Earth occultation catalog, Deep sample results 
(Harmon et al.~\citeauthor{har04}). 
$^{\rm (c)}$ The first COMPTEL source catalogue (Sch\"onfelder et~al.~\citeauthor{sch00}).
$^{\rm (d)}$ The third EGRET catalog of high-energy $\gamma$-ray sources 
(Hartman et al.~\citeauthor{har99}).
$^{\rm (e)}$ S: SIGMA instrument onboard GRANAT satellite; O: OSSE;
T: TeV source.
}
}
\end{table}

The top part of the
table is reserved for HMXBs, while the bottom part contains
those of low mass. In the second column of Table~\ref{detections} we list 
their flux (count/s) and error or upper limit in the energy range of
40$-$100~keV obtained with the IBIS $\gamma$-ray imager on board INTEGRAL, covering 
the first year of data (Bird et al.~\citeauthor{bir04}). 

The Burst and Transient Source Experiment (BATSE), aboard the CGRO, monitored the high energy sky using the Earth
occultation technique (EOT). A compilation of BATSE EOT observations has been
published recently (Harmon et al.~\citeauthor{har04}). From this catalogue we have
selected the data on microquasars in the energy range 160$-$430~keV in mCrab units
and is listed in the third column.

Among the
sources detected by the instrument COMPTEL (Sch\"onfelder et al.~\citeauthor{sch00}), 
also aboard the CGRO, there is the 
microquasar Cygnus~X-1, as well as two
sources, GRO~J1823$-$12 and GRO~J0241+6119, possibly associated with two other
microquasars. See fourth column in Table~\ref{detections}.

\section{Microquasars associated with EGRET sources}

According to the quasar-microquasar analogy (Mirabel \& Rodr\'{\i}guez
\citeauthor{mir99}), one could also expect the jets in microquasars to
be GeV emitters. Several models aimed to predict the high-energy
$\gamma$-ray emission from microquasars have been developed during the
last years. A general description of such models can be found in
Romero~(\citeauthor{rom04b}). Up to now, there are two HMXB
microquasars, LS~5039 and LS~I~+61~303, that are associated with two
EGRET sources.

\subsubsection{LS~5039 / 3EG~J1824$-$1514}

The discovery of the microquasar LS~5039, and its possible association with a
high-energy $\gamma$-ray source ($E>$100~MeV), provides observational evidence
that microquasars could also be sources of high-energy $\gamma$-rays (Paredes
et~al.~\citeauthor{par00}). It is important to point out that this was the first time
that an association between a microquasar and a high-energy $\gamma$-ray
source was reported. This finding opened up the possibility that other
unidentified EGRET sources could also be microquasars. LS~5039 is the only
X-ray source from the bright ROSAT catalogue whose position is consistent with
the high energy $\gamma$-ray source 3EG~J1824$-$1514. LS~5039 is also the only
object simultaneously detected in X-rays and radio 
which displays bipolar radio jets at sub-arcsecond scales.
New observations conducted with the EVN and MERLIN
confirm the presence of an asymmetric two-sided jet reaching up to
$\sim$1000~AU on the longest jet arm 
(Paredes et al.~\citeauthor{par02}, Rib\'o~\citeauthor{rib02}). 

\begin{figure}[] 
\vspace{6cm}
\includegraphics{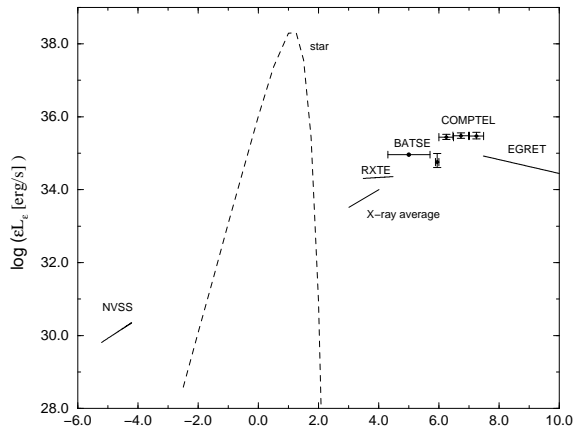}
\caption[]{Observed spectral energy distribution of LS~5039.}
\label{ls5039}
\end{figure}

Recently, Collmar~(\citeauthor{col03}) has reported the detection of an unidentified
$\gamma$-ray source, GRO~J1823$-$12, at galactic coordinates
($l$=17.5$^{\circ}$, b=$-$0.5$^{\circ}$) by the COMPTEL experiment. This source is among the strongest COMPTEL sources.
The source region, detected at a high significance level, contains several
possible counterparts, LS~5039 being one of them. It is also worth noting that
BATSE has detected this source at soft $\gamma$-rays (see
Table~\ref{detections}). Taking into account these observational evidences,
from radio to high-energy $\gamma$-rays, LS~5039 appears to be a very likely
counterpart of the EGRET source 3EG~J1824$-$1514.
Figure~\ref{ls5039} shows the observed spectral energy distribution of LS~5039.

\subsubsection{LS~I~+61~303 / 3EG~J0241+6103}

After the discovery of relativistic jets in LS~I~+61~303, this source has been classified as a new microquasar (Massi
et~al.~\citeauthor{mas01}, Massi et~al.~\citeauthor{mas04}). This object has also been
proposed to be associated with the $\gamma$-ray source 2CG~135+01
(=3EG~J0241+6103) (Gregory \& Taylor \citeauthor{gre78}, Kniffen
et~al.~\citeauthor{kni97}). Although the broadband 1~keV--100~MeV spectrum of
LS~I~+61~303 remains uncertain, because OSSE and COMPTEL observations were
likely dominated by the quasar QSO~0241+622 emission, the EGRET angular
resolution is high enough to exclude this quasar as the source of the
high-energy $\gamma$-ray emission (Harrison et~al.~\citeauthor{har00}).  BATSE marginally
detected the source, the quasar also being excluded as the origin of this
emission (see Table~\ref{detections}). 

Recently, Massi~(\citeauthor{mas04}) has carried out a timing analysis of pointed
EGRET observations (Tavani et al.~\citeauthor{tav98})
suggesting a period of 27.4$\pm$7.2 days, in agreement with the orbital period
of this binary system, of 26.496 days. This result, if confirmed, would
clearly support the association of LS~I~+61~303 with 3EG~J0241+6103.

This microquasar also seems to be a fast precessing system. MERLIN images
obtained in two consecutive days show a change in the
direction of the jets of about 50$^{\circ}$ that has been interpreted as a
fast precession of the system (Massi et al.~\citeauthor{masal04}). If this is
confirmed, it could solve the puzzling VLBI structures observed so far, as
well as the short term variability of the associated $\gamma$-ray source
3EG~J0241+6103. 

Up to now, the only existing radial velocity curve of LS~I~+61~303 was that
obtained by Hutchings and Crampton~(\citeauthor{hut81}). Recently, after a
spectroscopic campaign, an improved estimation of the orbital parameters has
been obtained (Casares et~al.~\citeauthor{cas04}). Here, we will just mention the
new high eccentricity (e=0.72$\pm$0.15) and the periastron orbital phase at
$\sim$0.2. These values are a key information for any interpretation of the
data obtained at any wavelength.

\begin{figure} 
\vspace{6cm}
\includegraphics{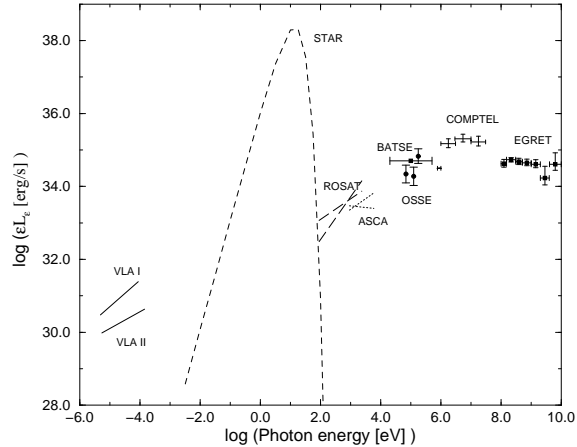}
\caption[]{Observed spectral energy distribution of LS~I~+61~303.}
\label{lsi61303}
\end{figure}

Hall et al.~(\citeauthor{hal03}) gave upper limits on the emission associated to
LS~I~+61~303 / 3EG~J0241+6103 at very high-energy $\gamma$-rays from
observations performed by the Cherenkov telescope Whipple. Several models have 
been proposed to explore the high
energy emission of this source (e.g. Taylor et al.~\citeauthor{tay96},
Punsly~\citeauthor{pun99}, Harrison et al.~\citeauthor{har00}, 
Leahy~\citeauthor{lea04}). The
most recent theoretical work has been presented by Bosch-Ramon \&
Paredes~(\citeauthor{bos04b}), who explore with a  detailed numerical model if this
system can both produce the emission and present the variability detected by
EGRET ($>$100~MeV). 
Figure~\ref{lsi61303} shows the observed spectral energy distribution of the microquasar LS~I~+61~303.

\begin{figure} 
\vspace{6cm}
\includegraphics{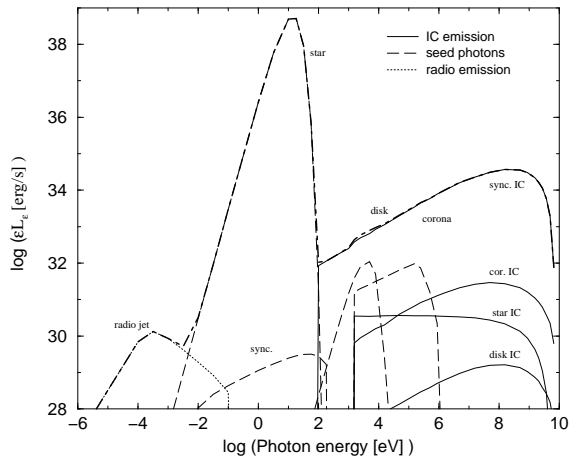}
\caption[]{Spectral energy distribution model of a high mass microquasar (Bosch-Ramon et~al. \citeauthor{bos04}).}
\label{model}
\end{figure}

Bosch-Ramon et~al.~(\citeauthor{bos04}) developed a detailed numerical model that includes both external and SSC scattering. The computed spectral energy distribution of a EGRET source high-mass microquasar is presented in Figure~\ref{model}. Looking at the Figures~\ref{ls5039}~and~\ref{lsi61303}, and comparing them with Figure~\ref{model}, it is seen how the IC jet scenario reproduces fairly well the data, giving further suport to the proposal of microquasars as $\gamma$-ray sources.


\section{VHE $\gamma$-ray sources}

The very high energy sky map contains a reduced number of sources. The number
of confirmed and probable catalogued sources presently amounts to fourteen (6 AGN, 3
pulsar wind nebulae, 3 supernova remnants, 1 starburst galaxy, and 1 unknown)
(Ong~\citeauthor{ong03}). Some microquasars have been observed in the energy range
of TeV $\gamma$-rays with the imaging atmospheric Cherenkov telescopes, but
none of them has been detected with high confidence yet. Historically,
Cygnus~X-3 was widely observed with the first generation of TeV instruments. 
Some groups claimed that they had detected Cygnus~X-3 (Chadwick
et~al.~\citeauthor{cha85}) whereas other groups failed to detect it (O'Flaherty 
et~al.~\citeauthor{ofl92}). As the claimed detections have not been confirmed, and
the instrumentation at this epoch was limited, these results have not been
considered as positive detections by the astronomical community. The HEGRA
experiment detected a flux of the order of 0.25~Crab from GRS~1915+105 during
the period May-July 1996 when the source was in an active state (Aharonian \&
Heinzelmann~\citeauthor{aha98}). This source has also been observed with Whipple,
obtaining a 3.1$\sigma$ significance (Rovero et~al.~\citeauthor{rov02}). More
recently, an upper-limit of 0.35 Crab above 400~GeV has been quoted for 
GRS~1915+105 (Horan \& Weekes~\citeauthor{hor03}). LS~I~+61~303 was observed too, but was not detected in the TeV energy range (see Section 4).

\acknowledgements
I acknowledge partial support by DGI of the Ministerio de Ciencia y
Tecnolog\'{\i}a (Spain) under grant AYA2001-3092, as well as partial support
by the European Regional Development Fund (ERDF/FEDER).



\end{article}
\end{document}